\documentclass[english,10pt]{article}
\usepackage[T1]{fontenc}
\usepackage[latin9]{inputenc}
\usepackage{calc}
\usepackage{amsmath}
\usepackage{graphicx}

\makeatletter
\@ifundefined{date}{}{\date{}}

\usepackage{pstricks, pst-node, pst-plot, pst-circ}
\usepackage{moredefs}

\usepackage{psfrag}

\usepackage{spconf,amsmath,graphicx}

\newcommand{\HP}{\text{H\hspace{-0.25mm}P}}
\newcommand{\LP}{\text{L\hspace{-0.25mm}P}}
\newcommand{\sigf}{f}

\newcommand{\fig}{Figure}
\newcommand{\tab}{Table}

\newcommand{\refsize}{\small}

\newcommand{\vspaceREF}{}

\makeatother

\usepackage{babel}
\begin{document}
\sloppy

\name{Wolfgang Schnurrer, Jürgen Seiler, Eugen Wige, and André Kaup}

\address{Multimedia Communications and Signal Processing\\
University of Erlangen-Nuremberg, Cauerstr. 7, 91058 Erlangen, Germany\\
Email: \{schnurrer, seiler, wige, kaup\}@lnt.de}


\title{Analysis of Displacement Compensation Methods for Wavelet Lifting
of  Medical 3-D Thorax CT Volume Data}

\maketitle


\begin{abstract}
A huge advantage of the wavelet transform in image and video compression
is its scalability. Wavelet-based coding of medical computed tomography
(CT) data becomes more and more popular. While much effort has been
spent on encoding of the wavelet coefficients, the extension of the
transform by a compensation method as in video coding has not gained
much attention so far. We will analyze two compensation methods for
medical CT data and compare the characteristics of the displacement
compensated wavelet transform with video data. We will show that for
thorax CT data the transform coding gain can be improved by a factor
of 2 and the quality of the lowpass band can be improved by 8~dB
in terms of PSNR compared to the original transform without compensation.
\end{abstract}

\section{Introduction}


Multi-dimensional volume data sets from CT or magnetic resonance
imaging (MRI) are often very large what challenges accessing, transmitting
or storing the data. A scalable representation yields the advantage
that a smaller version of the volume with lower resolution can be
used for previewing or browsing while the full resolution has to be
restored only when necessary. The wavelet transform can achieve a
scalable representation as the lowpass coefficients can be used as
downscaled version of the original signal.

Usually a multi-dimensional wavelet transform is realized by sequentially
applying a 1-D wavelet transform along the dimensions. Every transform
step decomposes the signal into two subbands containing the lowpass
and the highpass coefficients, respectively. As the lowpass coefficients
can be used as downscaled version of the original signal, a scalable
representation regarding the directions of the transform is achieved.
The wavelet transform can be factorized into a lifting representation
and introducing rounding operations results in an integer transform
\cite{calderbank1997}. That means that the original signal can be
restored from the transform coefficients without loss what makes the
integer wavelet transform interesting for medical applications. The
advantages of a scalable representation of multi-dimensional medical
volume data is analyzed in \cite{sanchez2010} with a special focus
on volume of interest coding.

\begin{figure}
\psfragscanon
\psfrag{Fil2}{$\sigf_{2i-2}$}
\psfrag{Fil1}{$\sigf_{2i-1}$}
\psfrag{Fi}{$\sigf_{2i}$}
\psfrag{Fir1}{$\sigf_{2i+1}$}
\psfrag{Fir2}{$\sigf_{2i+2}$}

\psfrag{Hil1}{$\HP_{i-1}$}
\psfrag{Hi}{$\HP_{i}$}

\psfrag{Lil1}{$\LP_{i-1}$}
\psfrag{Li}{$\LP_{i}$}
\psfrag{Lir1}{$\LP_{i+1}$}

\psfrag{MC}{MC}
\psfrag{IMC}{IMC}

\psfrag{pe}{$1$}
\psfrag{pv}{$\frac{1}{4}$}
\psfrag{mh}{$-\frac{1}{2}$}

\centering

\includegraphics[width=0.96\columnwidth]{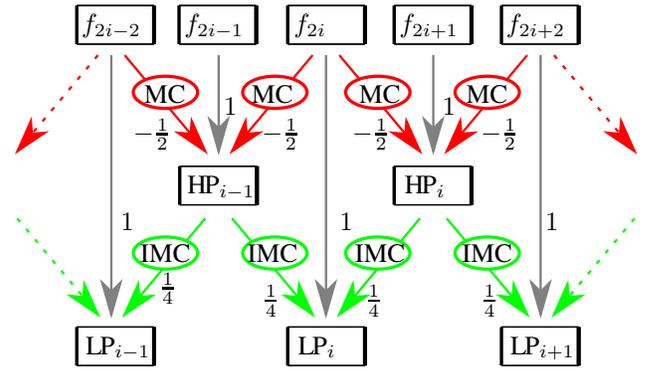}

\psfragscanoff

\protect\caption{\label{fig:LiftingStructure}Analysis step of the lifting structure
of the LeGall~5/3 wavelet with displacement compensation}
\end{figure}
With JPEG2000 a wavelet-based transform is already part of the DICOM
standard for coding medical images and multi-dimensional volumes can
be coded slice-wise with JPEG2000.

The disadvantage of the wavelet transform is the lowpass characteristic
of the lowpass band as it contains blurriness and motion artifacts.
That can reduce the quality and thus reduce the possible use of the
lowpass as downscaled version, e.g., for previewing purposes significantly.
A lot of effort has been spent on the coding of the wavelet coefficients
up to four dimensions for dynamic CT and MRI volumes \cite{zeng2003,lalgudi2005}.
In video coding it was already shown that the extension of the wavelet
lifting by motion compensation methods \cite{garbasTCSVT,Ohm1994}
improves the quality of the lowpass subband and thus improves the
scalability of the transform. In video sequences the displacement
between consecutive frames can mostly be described translatory and
thus modeled with block-based motion compensation methods.  Hence,
for video sequences a compensated wavelet transform along the time
axis is called motion compensated temporal filtering (MCTF) and for
a compensated wavelet transform along the view axis it is called disparity
compensated view filtering (DCVF), respectively \cite{garbasTCSVT}.
In medical CT volume data the displacement is rather describable as
deformation. The extension of the wavelet transform by a corresponding
displacement compensation has not been examined for medical volume
data so far. Nevertheless, block-based compensation methods were
gainfully used for coding medical volume data by using H.264/AVC \cite{sanchez2006,martin2008analysis}.

In this paper we examine whether a displacement compensation is advantageous
regarding the scalability in slice direction of medical CT volume
data. We therefor compare the behavior of a block-based and a mesh-based
displacement compensator. We examine whether the mesh-based method
is able to better compensate the more advanced displacement in medical
CT data. The characteristics of the compensated transform coefficients
are compared to the coefficients of the original wavelet transform
for medical CT volume data. To evaluate the performance of the displacement
compensated wavelet transform for medical CT data we show a comparison
to the MCTF approach of video data.

In Section~2 we review the extension of wavelet lifting by a compensation
step. We further introduce the metrics used for the evaluation. The
description of our simulation and results are given in Section~3.
Section~4 will conclude this study.

\section{Compensated Wavelet Lifting}

The filter representation of the wavelet transform can be factorized
into a lifting structure \cite{calderbank1997}. The analysis step
divides into a prediction step followed by an update step. We focus
on the LeGall~5/3 wavelet. In the prediction step the highpass coefficients
$\HP_{i}$ are computed to
\begin{equation}
\HP_{i}=\sigf_{2i+1}-\frac{1}{2}\left(\sigf_{2i}+\sigf_{2i+2}\right).\label{eq:H53-orig}
\end{equation}
The slices of the original volume are denoted by $\sigf_{i}$ where
$i$ denotes the slice index.  $2i$ denotes an even slice index
number and $2i+1$ denotes an odd slice index number. For a video
sequence, $f_{i}$ denotes the frame with index number $i$. In the
update step the lowpass coefficients $\LP_{i}$ are computed to 
\begin{equation}
\LP_{i}=\sigf_{2i}+\frac{1}{4}\left(\HP_{i-1}+\HP_{i}\right)\label{eq:L53-orig}
\end{equation}
using the results from the prediction step what leads to a significant
reduction of computational complexity compared to the . Omitting the
motion compensation (MC) and the inverse motion compensation (IMC)
for one moment this is also shown in \fig{}~\ref{fig:LiftingStructure}.
In the original lifting representation, the wavelet transform can
be easily extended by a compensation method. The compensation step
is denoted by MC in the prediction step in \fig{}~\ref{fig:LiftingStructure}.
For an equivalent to the wavelet transform an inversion of the compensation
is necessary in the update step, denoted by IMC in \fig{}~\ref{fig:LiftingStructure}.
We use the warping operator notation from \cite{garbasTCSVT} where
$\mathcal{W}_{ref\rightarrow cur}$ denotes the computation of a predictor
for the current slice with index $cur$ based on the reference slice
with index $ref$. The compensated coefficients of the highpass $\HP_{i}$
respectively the lowpass $\LP_{i}$ are then computed as follows.
\begin{equation}
\HP_{i}\hspace{-1mm}=\hspace{-1mm}\sigf_{2i+1}\hspace{-1mm}-\hspace{-1mm}\hspace{-0.4mm}\left\lfloor \frac{1}{2}\hspace{-0.4mm}\left(\mathcal{W}_{2i\rightarrow2i+1}\hspace{-1mm}\left(\sigf_{2i}\right)\hspace{-1mm}+\hspace{-1mm}\mathcal{W}_{2i+2\rightarrow2i+1}\hspace{-1mm}\left(\sigf_{2i+2}\right)\hspace{-0.4mm}\right)\hspace{-0.4mm}\right\rfloor \label{eq:H53}
\end{equation}
\begin{equation}
\LP_{i}\hspace{-1mm}=\hspace{-1mm}\sigf_{2i}\hspace{-1mm}+\hspace{-1mm}\left\lfloor \frac{1}{4}\left(\mathcal{W}_{2i-1\rightarrow2i}\hspace{-1mm}\left(\HP_{i-1}\right)\hspace{-1mm}+\hspace{-1mm}\mathcal{W}_{2i+1\rightarrow2i}\hspace{-1mm}\left(\HP_{i}\right)\right)\hspace{-0.4mm}\right\rfloor \label{eq:L53}
\end{equation}

As we also want to be able to reconstruct the volume without loss,
we use the integer implementation of the wavelet transform by introducing
a rounding operation \cite{calderbank1997} for the computation of
the fractal part.

For the computation of the highpass slices $\HP_{i}$, the predictors
from the previous reference slice $\sigf_{2i}$ and subsequent reference
slice $\sigf_{2i+2}$ are computed independently and are then combined.
The slice $\LP_{i}$ is computed from summation of $\sigf_{2i}$ with
the update term \eqref{eq:L53} and thus is quite similar to the original
slice $\sigf_{2i}$. Hence, we call $\sigf_{2i}$ of the original
signal the corresponding slice to slice $\LP_{i}$ of the lowpass
band.

\subsection{Displacement Compensation Methods}

The original lifting structure is fully equivalent to the filter representation
of the wavelet transform and will be denoted by an index `zero', e.g.,
$\HP_{\text{zero}}$.

The first compensation method is block-based. We use a fixed block
size and for every block we do a full search at integer positions
within the search range and choose the motion vector that minimizes
the sum of absolute differences \cite{fowler2000qccpack}. The block-based
compensation leads to a problem in the inversion step as pixels from
the reference slice can be multiple referenced or not referenced at
all. These pixels are called multiple connected and unconnected, respectively
\cite{Ohm1994}. Multiple connected pixels are averaged in the inversion
step. To fill the unconnected pixels we implemented the method in
\cite{bozinovic2005} where a nearest neighbor interpolation of the
motion vector field is proposed.

The block-based compensation without the hole filling in the update
step is denoted by the abbreviation `block' and block-based compensation
with the filling method is denoted by the abbreviation `block+fill'.

The second compensation method is mesh-based. A mesh grid is laid
over the reference slice and deformed to compute the predictor for
the current slice. The movement of the vertices is determined as follows.
For each vertex the neighboring pixels are used to get a block with
the vertex in the center \cite{fowler2000qccpack}. Then a block-based
search is performed as described in the paragraph above. For the vertices
on the border no movement is assumed. The advantage of this compensation
method is that it can be inverted without generating holes as for
the block-based compensation method.

The different displacement compensation methods are examined independently.
A multi-hypothesis approach that combines different displacement compensation
methods is not used.

\subsection{Metrics for Evaluation}

We want measure the energy compaction of the wavelet transform. We
therefor use the biorthogonal subband coding gain from \cite{Vaidyanathan1995}.
This measure for the energy compaction calculates the ratio of the
arithmetic and the geometric means of the subband variances and is
given by
\begin{equation}
G_{\text{SUB},\text{dcm}}=\frac{\sigma_{f}^{2}}{\sqrt{l_{\HP}^{2}\sigma_{\HP_{\text{dcm}}}^{2}}\cdot\sqrt{l_{\LP}^{2}\sigma_{\LP_{\text{dcm}}}^{2}}}.\label{eq:tcg}
\end{equation}
The variance $\sigma_{f}^{2}$ of the input volume $f$ is computed
by

\[
\sigma_{f}^{2}=\frac{1}{MNK}\sum_{mnk}\left(f_{k}\left[m,n\right]-\mu_{f}\right)^{2}
\]
where $\mu_{f}=\frac{1}{MNK}\sum_{mnk}f_{k}[m,n]$ is the mean of
the volume $f$, $K$ is the number of slices, $M$ is the number
of rows and $N$ is the number of columns of the volume $f$. So $f_{k}\left[m,n\right]$
denotes the voxel in the $m$-th row and the $n$-th column of the
$k$-th slice of the volume $f$. $\sigma_{\HP_{\text{dcm}}}^{2}$
and $\sigma_{\LP_{\text{dcm}}}^{2}$ denote the variances of the highpass
coefficients and the lowpass coefficients, respectively. The weighting
factors $l_{\LP}^{2}=\left(\frac{1+2+6+2+1}{8}\right)^{2}\approx0.72$
and $l_{\HP}^{2}=\left(\frac{1+2+1}{2}\right)^{2}=1.5$ are the squared
$l^{2}$-norms of the LeGall~5/3 wavelet filter impulse responses.
A higher energy compaction in one subband leads to a decrease of the
denominator in \eqref{eq:tcg}, i.e., the higher the energy compaction
of the wavelet transform the higher the subband transform coding gain
$G_{\text{SUB},\text{dcm}}$.

For a second analysis we evaluate the characteristics of the lowpass
coefficients. We therefor use two different metrics that both evaluate
the similarity of the lowpass band $\LP_{i}$ to the corresponding
slices of the original volume $\sigf_{2i}$$ $. We assume that the
higher the similarity, the less artifacts are contained in the lowpass
band. This means that the lowpass band is more usable as downscaled
version of the original volume.

With the first metric we evaluate the mean squared error (MSE) which
is an averaging similarity metric. Therefor we compute the MSE between
\begin{figure*}
\includegraphics[width=0.245\textwidth]{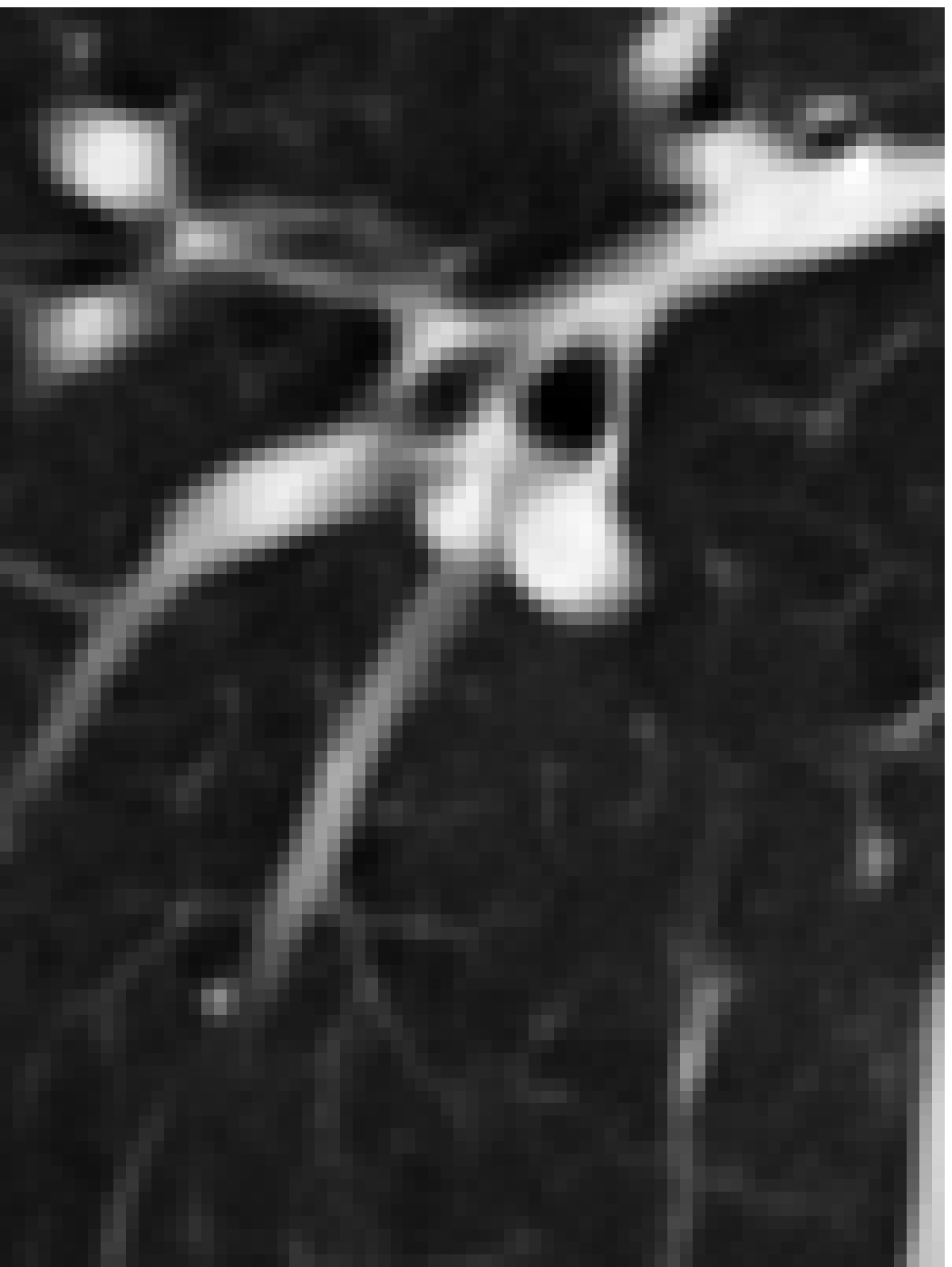}\hfill{}\includegraphics[width=0.245\textwidth]{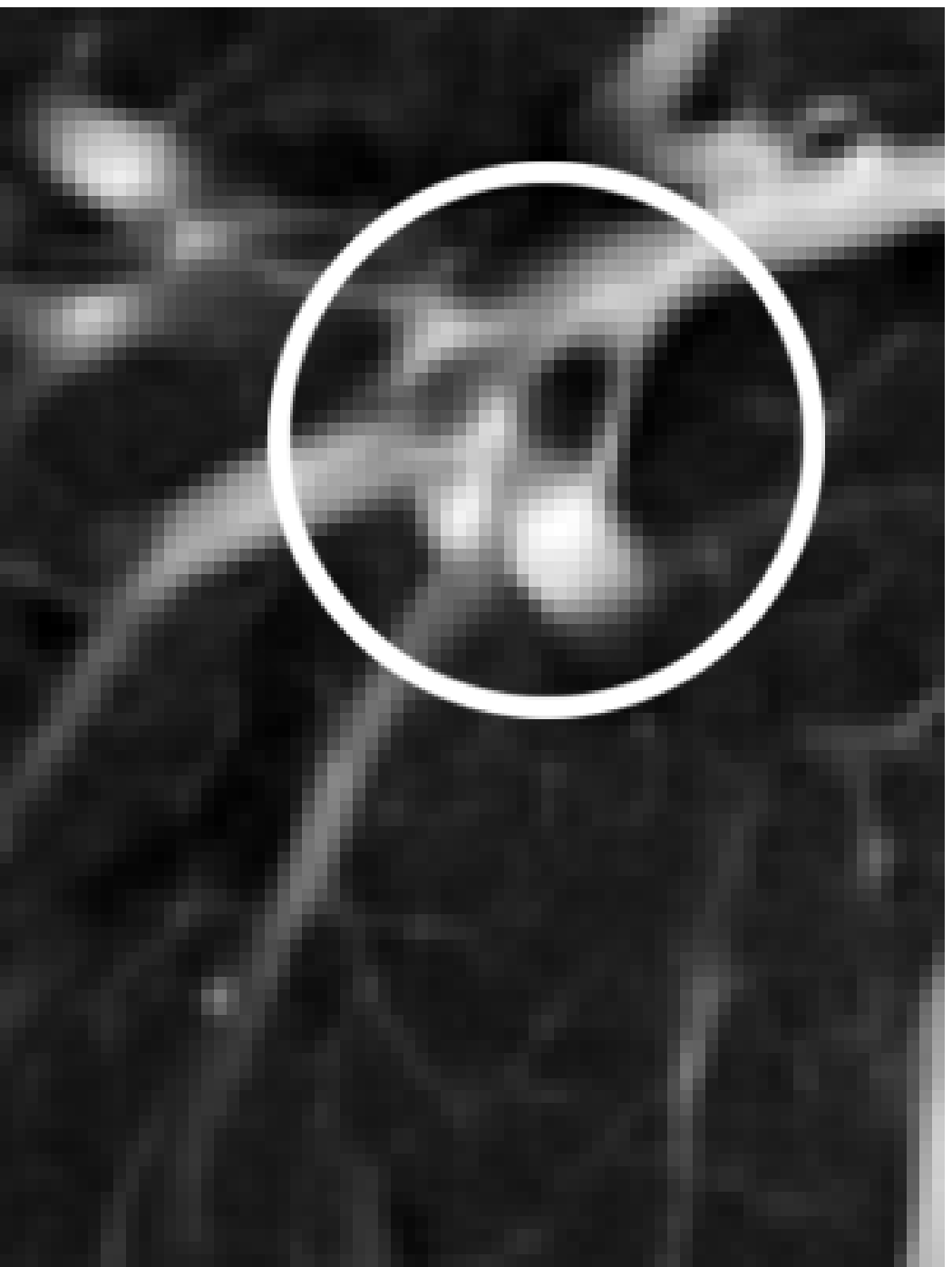}\hfill{}\includegraphics[width=0.245\textwidth]{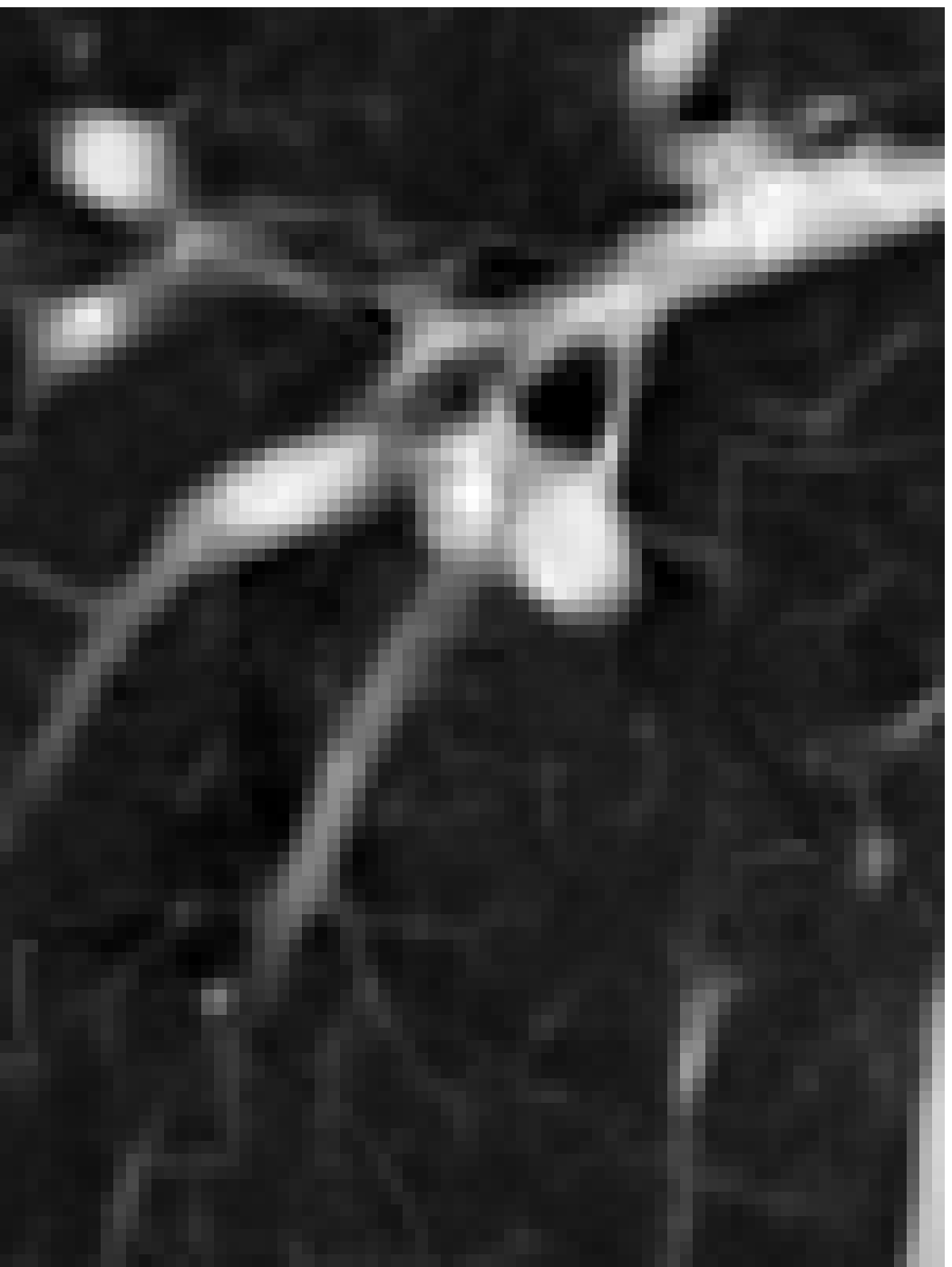}\hfill{}\includegraphics[width=0.245\textwidth]{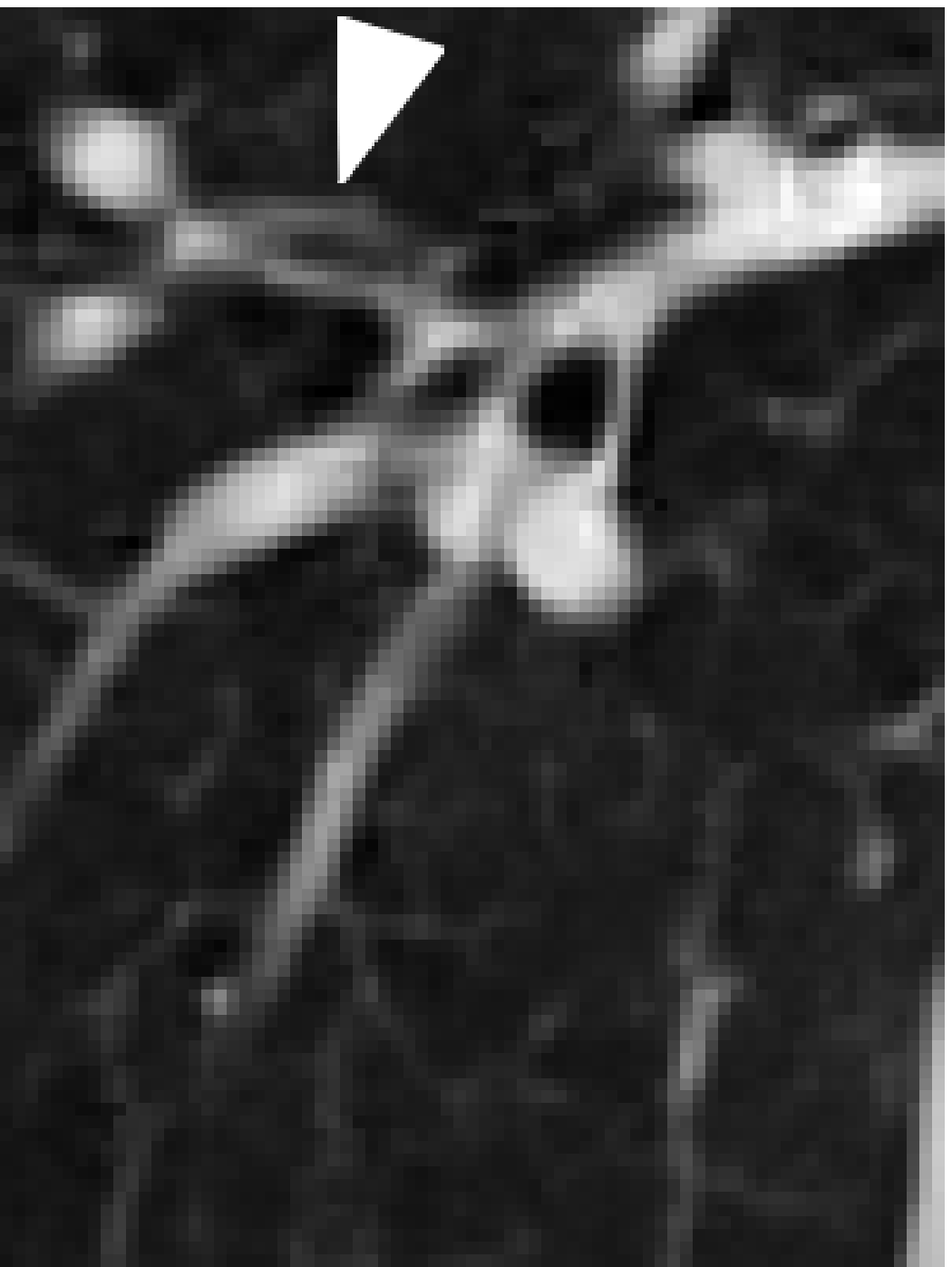}

\hspace{0.9cm}(a) original slice\hspace{1.8cm}(b) original  transform\hspace{1cm}(c)
mesh-based compensation\hspace{0.3cm}(d) block-based compensation

\hspace{5.8cm}(blurred)\hspace{3cm}(sharp)\hspace{3.5cm}(more details)

\vspace{-0.2cm}
\protect\caption{\label{fig:visualresults}Zoom in one slice of the lowpass band of
\textit{thorax2} for visual comparison, from left to right: (a) original
corresponding slice, (b) original wavelet transform without compensation,
(c) mesh-based compensation and (d) block-based compensation with
filling. The blur of the original uncompensated transform can be seen
especially in the marked circle. The block-based compensation method
is able to conserve more details as can be seen in the area marked
with an arrow in the right image.}
\end{figure*}
 the lowpass band and the corresponding original slices. For the original
transform without a compensation method, the MSE computes to
\begin{equation}
\text{MSE}\hspace{-0.5mm}\left(\LP_{\text{zero}},\hspace{-1mm}\sigf\right)\hspace{-1mm}=\hspace{-1mm}\frac{1}{M\hspace{-0.5mm}N\hspace{-0.5mm}K}\hspace{-1mm}\sum_{m,n,k}\hspace{-1mm}\hspace{-1mm}\left|\LP_{\text{zero},k}\hspace{-1mm}\left[m,\hspace{-0.5mm}n\right]\hspace{-1mm}-\hspace{-1mm}\sigf_{2k}\hspace{-1mm}\left[m,\hspace{-0.5mm}n\right]\right|^{2}\hspace{-1mm}.
\end{equation}
The computation of the $\text{MSE}\left(\LP_{\text{block}},\sigf\right)$
for the block-based displacement compensation, $\text{MSE}\left(\LP_{\text{block+fill}},\sigf\right)$
for the block-based displacement compensation with interpolation of
the motion vector field and the $\text{MSE}\left(\LP_{\text{mesh}},\sigf\right)$
for mesh-based displacement compensation method is analogous.

For computing the peak signal to noise ratio (PSNR), the MSE is computed
relative to the maximum intensity value $I_{\text{max}}$. The PSNR
for the whole lowpass band of the original transform without compensation
computes to
\begin{equation}
\text{PSNR}\left(\LP_{\text{zero}},\sigf\right)=10\log_{10}\left(\frac{I_{\text{max}}^{2}}{\text{MSE}\left(\LP_{\text{zero}},\sigf\right)}\right)\:\text{ [dB]}.
\end{equation}
For a more exact analysis, the PSNR is also computed per slice. As
different input data can have a different bit depth and thus a different
maximum intensity, a direct comparison of the PSNR values is difficult.
Subtracting two PSNR values results in a metric that gives the change
of the MSE in dB. This results in the lowpass gain
\begin{eqnarray}
G_{\LP,\text{MSE}} & = & \text{PSNR}\left(\LP_{\text{dcm}},\sigf\right)-\text{PSNR}\left(\LP_{\text{zero}},\sigf\right)\\
 & = & 10\log_{10}\left(\frac{\text{MSE}\left(\LP_{\text{zero}},\sigf\right)}{\text{MSE}\left(\LP_{\text{dcm}},\sigf\right)}\right)\:\text{ [dB}]\nonumber 
\end{eqnarray}
that gives the gain in reduction of the MSE relative to the original
transform in dB.  

With the $\text{L}_{\infty}$-norm, the second metric calculates the
maximum absolute distance between the lowpass band $\LP_{i}$ and
the corresponding original frames $\sigf_{2i}$. While the MSE is
an averaging metric the maximum absolute distance which is also desirable
to be low. For the original transform it computes to
\begin{equation}
\text{L}_{\infty}\left(\LP_{\text{zero}},\sigf\right)=\underset{\forall m,n,k}{\max}\left|\LP_{\text{zero},k}\left[m,n\right]-\sigf_{2k}\left[m,n\right]\right|.\label{eq:Linfty}
\end{equation}
The computation for `block', `block+fill' and `mesh' is again analogous.

\section{Results}

In our simulation we perform one wavelet decomposition step with and
without displacement compensation and compare the characteristics
of the resulting wavelet coefficients. We used a \textit{head} and
two \textit{thorax} 3-D CT data sets\footnote{The CT volume data sets were kindly provided by Prof. Dr. med. Dr.
rer. nat. Reinhard Loose from the Klinikum Nürnberg Nord.}. The CT data sets have a resolution of 512x512 with 32~slices (\textit{head}),
65~slices (\textit{thorax1}) and 67~slices (\textit{thorax2}) and
a bit depth of 12~bit per voxel. For comparison with video data we
took details from the HEVC test sequences in class~A, namely from
the sequences \textit{people} and \textit{traffic} with a resolution
of 512x512~pixels, all 150~frames. The CT data sets have an intensity
component only so we used the luminance component of the video sequences
that has a bit depth of 8~bit per pixel.

We used the implementation of the displacement compensation methods
in the QccPack library \cite{fowler2000qccpack}. As the library is
available in C, we implemented the method for filling the unconnected
pixels \cite{bozinovic2005} in C as well. For the block-based compensation
method we used a block size of 8x8~pixels and a search range of 8~pixels.
For the mesh-based method we also used a grid size of 8~pixels. The
blocks for the motion search have a size of 7x7~pixels and the search
range was set to 8~pixels. As our main intention was to examine
the properties of the extension of the wavelet transform for medical
volume data, we did not optimize our code for speed so far. The complexity
of the displacement estimation is comparable for the block-based and
the mesh-based approach as the movement of the vertices is determined
by block-matching. But the compensation step is a lot more complex
for the mesh-based method as an interpolation is needed for the values
in the deformed mesh.

\begin{figure*}[!p]

\input{fig/30-2--LpsnrF_norm_log_1on1.tex}\hspace{0.5cm}\input{fig/30-2--LmaxabsF_1on1.tex}

\hfill{}(a)\hfill{}\hfill{}(b)\hfill{}

\vspace{0.2cm}

\psfragscanon
\psfrag{m30}{zero}
\psfrag{m61}{block}
\psfrag{m60}{block+fill}
\psfrag{m100}{mesh}
\psfrag{m110}{weinlich}
\psfrag{m121}{block+den}
\psfrag{m120}{block+fill+den}

\fbox{\begin{minipage}[t]{0.97\linewidth}%
Legend: \hspace{0.5cm}\includegraphics[width=0.8\textwidth]{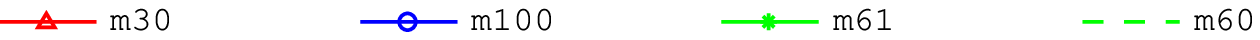}%
\end{minipage}}

\psfragscanoff

\vspace{0.2cm}

%
%
%
\providelength{\AxesLineWidth}       \setlength{\AxesLineWidth}{0.5pt}%
\providelength{\plotwidth}           \setlength{\plotwidth}{7.5cm}
\providelength{\LineWidth}           \setlength{\LineWidth}{0.7pt}%
\providelength{\MarkerSize}          \setlength{\MarkerSize}{4pt}%
\newrgbcolor{GridColor}{0.8 0.8 0.8}%
%
\psset{xunit=0.028571\plotwidth,yunit=0.049294\plotwidth}%
\begin{pspicture}(-3.500000,40.106612)(35.046667,59.352420)%


\psline[linewidth=\AxesLineWidth,linecolor=GridColor](0.000000,42.000000)(0.000000,42.243436)
\psline[linewidth=\AxesLineWidth,linecolor=GridColor](5.000000,42.000000)(5.000000,42.243436)
\psline[linewidth=\AxesLineWidth,linecolor=GridColor](10.000000,42.000000)(10.000000,42.243436)
\psline[linewidth=\AxesLineWidth,linecolor=GridColor](15.000000,42.000000)(15.000000,42.243436)
\psline[linewidth=\AxesLineWidth,linecolor=GridColor](20.000000,42.000000)(20.000000,42.243436)
\psline[linewidth=\AxesLineWidth,linecolor=GridColor](25.000000,42.000000)(25.000000,42.243436)
\psline[linewidth=\AxesLineWidth,linecolor=GridColor](30.000000,42.000000)(30.000000,42.243436)
\psline[linewidth=\AxesLineWidth,linecolor=GridColor](35.000000,42.000000)(35.000000,42.243436)
\psline[linewidth=\AxesLineWidth,linecolor=GridColor](0.000000,42.000000)(0.420000,42.000000)
\psline[linewidth=\AxesLineWidth,linecolor=GridColor](0.000000,44.000000)(0.420000,44.000000)
\psline[linewidth=\AxesLineWidth,linecolor=GridColor](0.000000,46.000000)(0.420000,46.000000)
\psline[linewidth=\AxesLineWidth,linecolor=GridColor](0.000000,48.000000)(0.420000,48.000000)
\psline[linewidth=\AxesLineWidth,linecolor=GridColor](0.000000,50.000000)(0.420000,50.000000)
\psline[linewidth=\AxesLineWidth,linecolor=GridColor](0.000000,52.000000)(0.420000,52.000000)
\psline[linewidth=\AxesLineWidth,linecolor=GridColor](0.000000,54.000000)(0.420000,54.000000)
\psline[linewidth=\AxesLineWidth,linecolor=GridColor](0.000000,56.000000)(0.420000,56.000000)
\psline[linewidth=\AxesLineWidth,linecolor=GridColor](0.000000,58.000000)(0.420000,58.000000)

{ \footnotesize 
\rput[t](0.000000,41.756564){$0$}
\rput[t](5.000000,41.756564){$5$}
\rput[t](10.000000,41.756564){$10$}
\rput[t](15.000000,41.756564){$15$}
\rput[t](20.000000,41.756564){$20$}
\rput[t](25.000000,41.756564){$25$}
\rput[t](30.000000,41.756564){$30$}
\rput[t](35.000000,41.756564){$35$}
\rput[r](-0.420000,42.000000){$42$}
\rput[r](-0.420000,44.000000){$44$}
\rput[r](-0.420000,46.000000){$46$}
\rput[r](-0.420000,48.000000){$48$}
\rput[r](-0.420000,50.000000){$50$}
\rput[r](-0.420000,52.000000){$52$}
\rput[r](-0.420000,54.000000){$54$}
\rput[r](-0.420000,56.000000){$56$}
\rput[r](-0.420000,58.000000){$58$}
} 

\psline[linewidth=\AxesLineWidth](0.000000,42.000000)(35.000000,42.000000)
\psline[linewidth=\AxesLineWidth](0.000000,42.000000)(0.000000,58.000000)

{ \small 
\rput[b](17.500000,40.106612){
\begin{tabular}{c}
index $i$ of lowpass slice $\LP_i$\\
\end{tabular}
}

\rput[t]{90}(-3.500000,50.000000){
\begin{tabular}{c}
$\text{PSNR} (\LP_{\text{dcm},i},\sigf_{2i})$ per slice in dB\\
\end{tabular}
}

\rput[t](17.500000,59.352420){
\begin{tabular}{c}
$$\textit{thorax2} lowpass: $\text{PSNR} (\LP_{\text{dcm},i},\sigf_{2i})$ per slice\\
\end{tabular}
}
} 

\newrgbcolor{color183.0217}{1  0  0}
\psline[plotstyle=line,linejoin=1,showpoints=true,dotstyle=Btriangle,dotsize=\MarkerSize,linestyle=solid,linewidth=\LineWidth,linecolor=color183.0217]
(2.000000,43.150880)(3.000000,43.596433)(4.000000,44.136780)(5.000000,44.069887)(6.000000,44.068691)
(7.000000,44.371791)(8.000000,44.526283)(9.000000,44.659712)(10.000000,44.610854)(11.000000,44.072395)
(12.000000,43.781322)(13.000000,43.713891)(14.000000,44.168740)(15.000000,44.360918)(16.000000,44.665842)
(17.000000,44.513962)(18.000000,44.411598)(19.000000,43.476556)(20.000000,43.529107)(21.000000,44.792471)
(22.000000,44.690440)(23.000000,44.377204)(24.000000,44.685097)(25.000000,45.805896)(26.000000,45.819537)
(27.000000,46.238275)(28.000000,46.037883)(29.000000,46.410349)(30.000000,47.025718)(31.000000,47.257285)
(32.000000,48.034262)(33.000000,48.917761)

\newrgbcolor{color184.0212}{0  1  0}
\psline[plotstyle=line,linejoin=1,showpoints=true,dotstyle=Basterisk,dotsize=\MarkerSize,linestyle=solid,linewidth=\LineWidth,linecolor=color184.0212]
(2.000000,52.694149)(3.000000,54.165239)(4.000000,55.267013)(5.000000,55.001061)(6.000000,54.700201)
(7.000000,54.263252)(8.000000,53.820635)(9.000000,53.826392)(10.000000,53.340448)(11.000000,53.169456)
(12.000000,52.013526)(13.000000,51.828576)(14.000000,52.162722)(15.000000,53.151092)(16.000000,53.824240)
(17.000000,53.353419)(18.000000,53.058321)(19.000000,52.492985)(20.000000,51.501770)(21.000000,53.897715)
(22.000000,52.831999)(23.000000,51.532429)(24.000000,53.221325)(25.000000,55.452577)(26.000000,54.372074)
(27.000000,54.568498)(28.000000,55.229784)(29.000000,55.871870)(30.000000,55.657967)(31.000000,55.583035)
(32.000000,55.387783)(33.000000,56.047813)

\newrgbcolor{color185.0212}{0  1  0}
\psline[plotstyle=line,linejoin=1,linestyle=dashed,linewidth=\LineWidth,linecolor=color185.0212]
(2.000000,50.189789)(3.000000,51.583392)(4.000000,52.938291)(5.000000,52.704892)(6.000000,52.522358)
(7.000000,52.131489)(8.000000,51.368267)(9.000000,51.358457)(10.000000,50.999714)(11.000000,50.875434)
(12.000000,49.551383)(13.000000,49.419495)(14.000000,49.877586)(15.000000,50.864592)(16.000000,51.332625)
(17.000000,50.772185)(18.000000,50.685859)(19.000000,49.829494)(20.000000,48.592419)(21.000000,51.476908)
(22.000000,50.218391)(23.000000,49.415959)(24.000000,50.512245)(25.000000,53.371356)(26.000000,51.851029)
(27.000000,52.258578)(28.000000,52.562399)(29.000000,52.935874)(30.000000,53.410992)(31.000000,53.102876)
(32.000000,53.135002)(33.000000,53.918703)

\newrgbcolor{color186.0212}{0  0  1}
\psline[plotstyle=line,linejoin=1,showpoints=true,dotstyle=Bo,dotsize=\MarkerSize,linestyle=solid,linewidth=\LineWidth,linecolor=color186.0212]
(2.000000,47.574358)(3.000000,48.867219)(4.000000,50.601765)(5.000000,49.959850)(6.000000,50.057979)
(7.000000,49.288417)(8.000000,49.221744)(9.000000,48.801170)(10.000000,48.652062)(11.000000,48.188918)
(12.000000,46.976441)(13.000000,47.136993)(14.000000,47.312553)(15.000000,48.138899)(16.000000,48.346403)
(17.000000,48.552914)(18.000000,48.314181)(19.000000,47.748643)(20.000000,46.133584)(21.000000,48.540008)
(22.000000,47.510722)(23.000000,47.665309)(24.000000,48.272661)(25.000000,50.959519)(26.000000,49.503636)
(27.000000,50.424007)(28.000000,49.838657)(29.000000,49.562630)(30.000000,51.131966)(31.000000,51.091164)
(32.000000,50.980539)(33.000000,51.793485)

\end{pspicture}%
\hspace{0.5cm}
%
%
%
\providelength{\AxesLineWidth}       \setlength{\AxesLineWidth}{0.5pt}%
\providelength{\plotwidth}           \setlength{\plotwidth}{7.5cm}
\providelength{\LineWidth}           \setlength{\LineWidth}{0.7pt}%
\providelength{\MarkerSize}          \setlength{\MarkerSize}{4pt}%
\newrgbcolor{GridColor}{0.8 0.8 0.8}%
%
\psset{xunit=0.028571\plotwidth,yunit=0.001972\plotwidth}%
\begin{pspicture}(-4.200000,52.665303)(35.046667,533.810498)%


\psline[linewidth=\AxesLineWidth,linecolor=GridColor](0.000000,100.000000)(0.000000,106.085890)
\psline[linewidth=\AxesLineWidth,linecolor=GridColor](5.000000,100.000000)(5.000000,106.085890)
\psline[linewidth=\AxesLineWidth,linecolor=GridColor](10.000000,100.000000)(10.000000,106.085890)
\psline[linewidth=\AxesLineWidth,linecolor=GridColor](15.000000,100.000000)(15.000000,106.085890)
\psline[linewidth=\AxesLineWidth,linecolor=GridColor](20.000000,100.000000)(20.000000,106.085890)
\psline[linewidth=\AxesLineWidth,linecolor=GridColor](25.000000,100.000000)(25.000000,106.085890)
\psline[linewidth=\AxesLineWidth,linecolor=GridColor](30.000000,100.000000)(30.000000,106.085890)
\psline[linewidth=\AxesLineWidth,linecolor=GridColor](35.000000,100.000000)(35.000000,106.085890)
\psline[linewidth=\AxesLineWidth,linecolor=GridColor](0.000000,100.000000)(0.420000,100.000000)
\psline[linewidth=\AxesLineWidth,linecolor=GridColor](0.000000,150.000000)(0.420000,150.000000)
\psline[linewidth=\AxesLineWidth,linecolor=GridColor](0.000000,200.000000)(0.420000,200.000000)
\psline[linewidth=\AxesLineWidth,linecolor=GridColor](0.000000,250.000000)(0.420000,250.000000)
\psline[linewidth=\AxesLineWidth,linecolor=GridColor](0.000000,300.000000)(0.420000,300.000000)
\psline[linewidth=\AxesLineWidth,linecolor=GridColor](0.000000,350.000000)(0.420000,350.000000)
\psline[linewidth=\AxesLineWidth,linecolor=GridColor](0.000000,400.000000)(0.420000,400.000000)
\psline[linewidth=\AxesLineWidth,linecolor=GridColor](0.000000,450.000000)(0.420000,450.000000)
\psline[linewidth=\AxesLineWidth,linecolor=GridColor](0.000000,500.000000)(0.420000,500.000000)

{ \footnotesize 
\rput[t](0.000000,93.914110){$0$}
\rput[t](5.000000,93.914110){$5$}
\rput[t](10.000000,93.914110){$10$}
\rput[t](15.000000,93.914110){$15$}
\rput[t](20.000000,93.914110){$20$}
\rput[t](25.000000,93.914110){$25$}
\rput[t](30.000000,93.914110){$30$}
\rput[t](35.000000,93.914110){$35$}
\rput[r](-0.420000,100.000000){$100$}
\rput[r](-0.420000,150.000000){$150$}
\rput[r](-0.420000,200.000000){$200$}
\rput[r](-0.420000,250.000000){$250$}
\rput[r](-0.420000,300.000000){$300$}
\rput[r](-0.420000,350.000000){$350$}
\rput[r](-0.420000,400.000000){$400$}
\rput[r](-0.420000,450.000000){$450$}
\rput[r](-0.420000,500.000000){$500$}
} 

\psline[linewidth=\AxesLineWidth](0.000000,100.000000)(35.000000,100.000000)
\psline[linewidth=\AxesLineWidth](0.000000,100.000000)(0.000000,500.000000)

{ \small 
\rput[b](17.500000,52.665303){
\begin{tabular}{c}
index $i$ of lowpass slice $\LP_i$\\
\end{tabular}
}

\rput[t]{90}(-4.200000,300.000000){
\begin{tabular}{c}
$\text{L}_{\infty} (\LP_{\text{dcm},i},\sigf_{2i})$\\
\end{tabular}
}

\rput[t](17.500000,533.810498){
\begin{tabular}{c}
$$\textit{thorax2} lowpass: $\text{L}_{\infty}$-norm per slice\\
\end{tabular}
}
} 

\newrgbcolor{color183.015}{1  0  0}
\psline[plotstyle=line,linejoin=1,showpoints=true,dotstyle=Btriangle,dotsize=\MarkerSize,linestyle=solid,linewidth=\LineWidth,linecolor=color183.015]
(2.000000,490.000000)(3.000000,335.000000)(4.000000,216.000000)(5.000000,228.000000)(6.000000,262.000000)
(7.000000,264.000000)(8.000000,292.000000)(9.000000,378.000000)(10.000000,423.000000)(11.000000,312.000000)
(12.000000,337.000000)(13.000000,370.000000)(14.000000,478.000000)(15.000000,305.000000)(16.000000,336.000000)
(17.000000,346.000000)(18.000000,346.000000)(19.000000,386.000000)(20.000000,293.000000)(21.000000,373.000000)
(22.000000,284.000000)(23.000000,298.000000)(24.000000,207.000000)(25.000000,188.000000)(26.000000,210.000000)
(27.000000,266.000000)(28.000000,268.000000)(29.000000,278.000000)(30.000000,290.000000)(31.000000,292.000000)
(32.000000,287.000000)(33.000000,288.000000)

\newrgbcolor{color184.0145}{0  1  0}
\psline[plotstyle=line,linejoin=1,showpoints=true,dotstyle=Basterisk,dotsize=\MarkerSize,linestyle=solid,linewidth=\LineWidth,linecolor=color184.0145]
(2.000000,281.000000)(3.000000,153.000000)(4.000000,146.000000)(5.000000,165.000000)(6.000000,261.000000)
(7.000000,255.000000)(8.000000,277.000000)(9.000000,227.000000)(10.000000,306.000000)(11.000000,347.000000)
(12.000000,261.000000)(13.000000,229.000000)(14.000000,354.000000)(15.000000,210.000000)(16.000000,283.000000)
(17.000000,192.000000)(18.000000,178.000000)(19.000000,185.000000)(20.000000,145.000000)(21.000000,212.000000)
(22.000000,193.000000)(23.000000,172.000000)(24.000000,213.000000)(25.000000,153.000000)(26.000000,150.000000)
(27.000000,205.000000)(28.000000,208.000000)(29.000000,216.000000)(30.000000,200.000000)(31.000000,188.000000)
(32.000000,192.000000)(33.000000,153.000000)

\newrgbcolor{color185.0145}{0  1  0}
\psline[plotstyle=line,linejoin=1,linestyle=dashed,linewidth=\LineWidth,linecolor=color185.0145]
(2.000000,353.000000)(3.000000,244.000000)(4.000000,149.000000)(5.000000,165.000000)(6.000000,264.000000)
(7.000000,255.000000)(8.000000,290.000000)(9.000000,300.000000)(10.000000,306.000000)(11.000000,347.000000)
(12.000000,296.000000)(13.000000,259.000000)(14.000000,357.000000)(15.000000,210.000000)(16.000000,283.000000)
(17.000000,229.000000)(18.000000,178.000000)(19.000000,249.000000)(20.000000,280.000000)(21.000000,226.000000)
(22.000000,193.000000)(23.000000,196.000000)(24.000000,213.000000)(25.000000,180.000000)(26.000000,198.000000)
(27.000000,205.000000)(28.000000,208.000000)(29.000000,216.000000)(30.000000,232.000000)(31.000000,188.000000)
(32.000000,205.000000)(33.000000,188.000000)

\newrgbcolor{color186.0145}{0  0  1}
\psline[plotstyle=line,linejoin=1,showpoints=true,dotstyle=Bo,dotsize=\MarkerSize,linestyle=solid,linewidth=\LineWidth,linecolor=color186.0145]
(2.000000,437.000000)(3.000000,240.000000)(4.000000,220.000000)(5.000000,265.000000)(6.000000,262.000000)
(7.000000,297.000000)(8.000000,303.000000)(9.000000,373.000000)(10.000000,291.000000)(11.000000,257.000000)
(12.000000,299.000000)(13.000000,388.000000)(14.000000,408.000000)(15.000000,320.000000)(16.000000,266.000000)
(17.000000,283.000000)(18.000000,252.000000)(19.000000,247.000000)(20.000000,238.000000)(21.000000,273.000000)
(22.000000,252.000000)(23.000000,248.000000)(24.000000,211.000000)(25.000000,217.000000)(26.000000,316.000000)
(27.000000,206.000000)(28.000000,375.000000)(29.000000,364.000000)(30.000000,216.000000)(31.000000,186.000000)
(32.000000,243.000000)(33.000000,229.000000)

\end{pspicture}%

\hfill{}(c)\hfill{}\hfill{}(d)\hfill{}

\protect\caption{\label{fig:results-Lband-detailled}Detailed results for the sequence
\textit{people} in the first row (a), (b) and for the medical CT volume
\textit{thorax2} in the second row (c), (d). First column: PSNR of
the lowpass band and the corresponding original slices, second column:
the $\text{L}_{\infty}$-norm per slice of the lowpass band and the
corresponding original slices. The placeholder `dcm' stands for the
displacement compensation method.}
\end{figure*}

\begin{table}[!t]
\begin{tabular*}{1\linewidth}{@{\extracolsep{\fill}}|c||c|c|c|c|}
\hline
 & \multicolumn{4}{c|}{

subband transform coding gain $G_{\text{SUB},\text{dcm}}$}\tabularnewline
\hline
& zero & mesh & block & block+fill\tabularnewline

\hline
\hline

\textit{people} & 3.31 & 6.44 & \textbf{9.98} & 9.97 \tabularnewline

\hline

\textit{traffic} & 4.12 & 10.17 & \textbf{11.49} & 11.48 \tabularnewline

\hline

\textit{thorax1} & 5.45 & 7.96 & \textbf{13.08} & \textbf{13.08} \tabularnewline

\hline

\textit{thorax2} & 6.21 & 9.19 & \textbf{14.27} & 14.26 \tabularnewline

\hline

\textit{head} & 5.52 & 6.45 & \textbf{8.5} & 8.49 \tabularnewline

\hline
\end{tabular*}

\protect\caption{\label{tab:Results-TCG-values}Subband coding gain, `dcm' is a placeholder
for the displacement compensation methods the row below}
\vspace{-0.2cm}
\end{table}
\begin{table*}[!p]
\begin{tabular*}{1\linewidth}{@{\extracolsep{\fill}}|c||c|c|c|c||c|c|c||c|c|c|c|}
\hline
 & \multicolumn{4}{c||}{lowpass $\text{PSNR}\left(\LP_{\text{dcm}},f\right)$ in dB} & \multicolumn{3}{c||}{lowpass
gain $G_{\LP,\text{MSE}}$ in dB} & \multicolumn{4}{c|}{$\text{L}_{\infty}\left(\LP_{\text{dcm}},\sigf\right)$}\tabularnewline
\hline
& zero & mesh & block & block+fill & mesh & block & block+fill & zero & mesh & block & block+fill\tabularnewline

\hline
\hline

\textit{people} & 32.64 & 38.93 & \textbf{43.47} & 41.44 & 6.29 & \textbf{10.83} & 8.8 & 83 & 66 & \textbf{57} & \textbf{57}\tabularnewline


\textit{traffic} & 38.01 & 46.39 & \textbf{47.19} & 46.56 & 8.38 & \textbf{9.18} & 8.55 & 62 & 68 & \textbf{60} & \textbf{60}\tabularnewline


\textit{thorax1} & 44.01 & 47.59 & \textbf{52.71} & 50.27 & 3.58 & \textbf{8.7} & 6.26 & 378 & 567 & \textbf{318} & 339\tabularnewline


\textit{thorax2} & 44.75 & 48.76 & \textbf{53.65} & 51.21 & 4.01 & \textbf{8.9} & 6.46 & 490 & 437 & \textbf{354} & 357\tabularnewline


\textit{head} & 43.27 & 43.74 & \textbf{46.56} & 44.84 & 0.47 & \textbf{3.29} & 1.57 & \textbf{502} & 1006 & 754 & 754

\\ \hline
\end{tabular*}

\protect\caption{\label{tab:Results-Mean-values-Lband}Lowpass band results, PSNR on
the left, $G_{\LP,\text{MSE}}$ in the middle, and $ $$\text{L}_{\infty}\left(\LP_{\text{dcm}},\sigf\right)$
on the right. `dcm' is a placeholder for the displacement compensation
methods the row below}
\end{table*}
We perform a single wavelet decomposition step and then compare the
characteristics of the highpass and the lowpass coefficients. For
the video sequences the transform is performed along the time axis
and for the CT volumes along the slice axis. The simulation results
are summarized in \tab{}~\ref{tab:Results-TCG-values} and \tab{}~\ref{tab:Results-Mean-values-Lband}.
More detailed results are shown in \fig{}~\ref{fig:results-Lband-detailled}
where the metrics are evaluated and plotted per slice for the video
sequence \textit{people} and the medical CT volume \textit{thorax2}.

For visual comparison of the lowpass bands of the different approaches,
a zoom into one slice of the lowpass band of \textit{thorax2} is shown
in \fig{}~\ref{fig:visualresults}. Compared to the corresponding
original slice in \fig{}~\ref{fig:visualresults}~(a) the blurriness
of the lowpass band of the original transform can be seen in \fig{}~\ref{fig:visualresults}~(b).
The lowpass bands of the compensated transforms in \fig{}~\ref{fig:visualresults}~(c)
and (d) represent the structures sharper and more detailed.

For the quantitative measures we first compare the subband transform
coding gain reduction gain  $G_{\text{SUB},\text{dcm}}$. The results
are listed in \tab{}~\ref{tab:Results-TCG-values}. The interpolation
of the motion vector field (`block+fill') has only minor influence
on $G_{\text{SUB},\text{dcm}}$. For all examined data, the compensation
methods lead to a significant increase of the subband transform coding
gain and thus improve the energy compaction of the wavelet transform.
 Further, the block-based method is more capable of improving the
transform  for both kinds of data.  For both thorax CT data sets,
the transform gain can be improved by more than a factor of 2.

Second we compare the similarity of the lowpass band and the corresponding
original slices. The first columns of \tab{}~\ref{tab:Results-Mean-values-Lband}
contain the lowpass $\text{PSNR}\left(\LP_{\text{dcm}},f\right)$
in dB. The PSNR results suggest, that a compensation method is advantageous
for all examined data. The gain of the compensation methods relative
to the original transforms without compensation is listed in the columns
entitled by lowpass gain $G_{\LP,\text{MSE}}$ in dB.   For both
kinds of data, the lowpass bands are more similar to the original
slices for a compensated wavelet transform. While the gains for the
CT volume \textit{head} are lower, the gains for the displacement
compensated transform of the \textit{thorax} volumes are comparable
to the gains for the MCTF of the video sequences. The interpolation
of the motion vector field leads to a lower gain in MSE reduction
but still higher than for the mesh-based method. The more detailed
slice-wise results in \fig{}~\ref{fig:results-Lband-detailled}~(a)
show a very similar behavior of the compensation methods for the video
sequence \textit{people} compared to the results in \fig{}~\ref{fig:results-Lband-detailled}~(c)
for the CT volume \textit{thorax2}.

Third we compare the $\text{L}_{\infty}$-norm of the lowpass band
and the corresponding original slices. The higher values for the medical
CT volumes can be explained by the different bit depth. For all examined
data, the $\text{L}_{\infty}$-norm for the mesh-based method is higher
than for the block-based method. That means that the block-based method
can model the occurring displacement better. For the video sequences
the $\text{L}_{\infty}$-norm can be reduced by incorporating either
of the two compensation methods while the block-based method leads
to smaller values. For the CT volume \textit{head} the maximum absolute
distances are significantly higher for the coefficients resulting
from the displacement compensated transform. That shows that the examined
methods are not suitable to compensate the displacement in all kinds
of medical CT volumes in general and, e.g., for the CT volume \textit{head}
a proper displacement compensation method needs to be found. For the
two \textit{thorax} CT volumes the results of the block-based method
are similar to the video sequences though the reduction is smaller.
The more detailed results in \fig{}~\ref{fig:results-Lband-detailled}~(b)
and (d) also show a similar behavior  for the medical CT volume \textit{thorax2}.

\section{Conclusion}

For all examined types of data, the introduction of a displacement
compensation step into the lifting structure leads to an  increase
of the transform coding gain by more than a factor of 2 and thus a
better energy compaction for video data and \textit{thorax} CT volumes.
 In our simulations the block-based displacement compensation performed
better than the mesh-based method.

Further, we could show that a displacement compensation step can improve
the similarity of the lowpass band to the corresponding original also
for medical \textit{thorax} CT volumes by 8 dB in terms of PSNR. This
enlarges the quality and thus the usability of the lowpass band as
downscaled version of the original volume. A next step is to examine
the extension of the wavelet transform by a compensation step along
the time axis of dynamical medical volume data.

It is not clear whether the compensated lowpass coefficients can be
used directly  for diagnostic purposes. Nevertheless, a scalable
representation is advantageous for faster and more efficient browsing
in a huge data volume. This can be very advantageous,  e.g., in telemedicine.

Further work aims at a more detailed complexity analysis and an improvement
of the inversion of the block-based compensation.

\subsection*{Acknowledgment}

We gratefully acknowledge that this work has been supported by the
Deutsche Forschungsgemeinschaft (DFG) under contract number KA~926/4-1.

\end{document}